\documentclass[11pt]{article}
\immediate\write18{perl ../../scripts/bibgen2 \jobname.tex}
\usepackage{amssymb,amsmath,graphicx}
\newcommand{\be}{\begin{equation}}
\newcommand{\ee}{\end{equation}}
\newcommand{\bea}{\begin{eqnarray}\displaystyle}
\newcommand{\eea}{\end{eqnarray}}
\newcommand{\bdm}{\begin{displaymath}}
\newcommand{\edm}{\end{displaymath}}
\newcommand{\sectiono}[1]{\section{#1}\setcounter{equation}{0}}

\newcommand{\Tr}{\mathop{\rm Tr}\nolimits}

\def\bra#1{\langle #1 |}
\def\ket#1{|#1 \rangle}

\def\aver#1{\langle\, #1 \,\rangle}

\def\cN{\mathcal{N}}
\def\bartilde#1{\bar{\tilde #1}}

\begin{document}
{}~ \hfill\vbox{\hbox{arXiv:0807.1102} }\break \vskip 2.1cm

\centerline{\Large \bf Classification of $\cN=6$
superconformal theories of ABJM type}
\vspace*{8.0ex}

\centerline{\large \rm Martin Schnabl and Yuji Tachikawa}

\vspace*{8.0ex}

\centerline{\large \it School of Natural Sciences, Institute for Advanced
Study,} \centerline{\large \it Princeton, NJ 08540, USA} \vspace*{2.0ex}
\centerline{E-mail: {\tt schnabl, yujitach at ias.edu}}

\vspace*{6.0ex}

\centerline{\bf Abstract}
\bigskip

Studying the supersymmetry enhancement mechanism of Aharony, Bergman, Jafferis
and Maldacena, we find a simple condition on the gauge group generators for the
matter fields. We analyze all possible compact Lie groups and their
representations. The only allowed gauge groups leading to the manifest ${\cal
N}=6$ supersymmetry are, up to discrete quotients, $SU(n) \times U(1)$, $Sp(n)
\times U(1)$, $SU(n) \times SU(n)$, and $SU(n) \times SU(m) \times U(1)$ with
possibly additional $U(1)$'s. Matter representations are restricted to be the (bi)fundamentals.
As a byproduct we obtain another proof of the
complete classification of the three algebras considered by Bagger and Lambert.

\vfill \eject

\baselineskip=16pt


\sectiono{Introduction and Conclusions}
\label{s_intro}

Recently superconformal Chern-Simons theories in three dimensions \cite{sch} 
have attracted renewed interest,
after the discovery of the $\cN=8$  Lagrangian in \cite{BL1,BL2,G}.
Soon afterwards $\cN=4$  and $\cN=6$ Lagrangians were constructed by
\cite{GW,H1} and \cite{ABJM,K} respectively.
These theories have an interesting property that the closure of the supersymmetry
requires particular combinations of the gauge group and the matter content,
whereas there is no such restriction  for $\cN\le 3$.

An illuminating way to understand this enhancement of supersymmetry for
particular matter contents and gauge groups was presented in  \cite{ABJM},
which we review shortly. The arguments in \cite{ABJM} were for the enhancement
to $\cN=6$ and $8$, but the generalization to the case $\cN=4,5$ is also
straightforward, as will be briefly mentioned here. We follow the notation of \cite{YG}.

Firstly, recall that $\cN=2$  theories can be formulated for any gauge group
$G$ and chiral superfields in any representation, with arbitrary
superpotential. An $\cN=3$ superconformal theory in the $\cN=2$ language has
superfields $Q^A$ ($A=1,\ldots,2n$) in the pseudoreal representation\footnote{
Matter contents might be easier to understand from the viewpoint of $d=3$
$\cN=4$ theory, which has almost the same structure as $d=4$ $\cN=2$ theory. A
$d=4$, $\cN=2$ vector multiplet decomposes into a $d=4$, $\cN=1$ vector
multiplet and a chiral multiplet, both in the adjoint representation. The
flavor symmetry of $n$ free hypermultiplets $Q_A$ is $Sp(n)$, therefore the
gauge group $G$ needs to act on the hypermultiplet as a $2n$-dimensional
pseudoreal representation $R_0$. This is usually referred to as a
half-hypermultiplet in the representation $R_0$. When $R_0$ is a direct sum
$R\oplus \bar R$ of a complex representation $R$ and its conjugate, $Q_A$
decomposes into two chiral multiplets $Q_a$ $(a=1,\ldots,n)$ in $R$ and $\tilde
Q_{\bar a}$ ($\bar a=1,\ldots,n$) in $\bar R$. This is the full hypermultiplet
in the representation $R$. In $d=4$ the half-hypermultiplet is often afflicted
with global anomaly, but this problem does not exist in $d=3$.} $R_0$,  one
auxiliary chiral superfield $\Phi_a$ in the adjoint, and the superpotential
coupling $\sim Q^A Q^B T^a_{AB} \Phi_a$. Here $A,B$ are the indices of the
representation $R_0$, and $a$ is the adjoint index. Note that $T^a_{AB}$ is
symmetric in $A$ and $B$. The important point is that the $\cN=3$ theory has R-symmetry
$SO(3)_R=SU(2)_R$ which rotates  as doublet the lowest component of the chiral
superfield $Q$ and of its conjugate, anti-chiral superfield $Q^\dagger$.
Enhancement of the supersymmetry is due to the interplay of this $SU(2)_R$
symmetry, which is not manifest in $\cN=2$ formalism, and  an enhancement of
the flavor symmetry of the $\cN=2$ superpotential discussed below.

Now, let us eliminate the auxiliary fields $\Phi$.
It gives the superpotential \begin{equation}
W=f_{ABCD}Q^AQ^BQ^CQ^D,
\end{equation} where
\be\label{f-def}
f_{ABCD} = K_{pq} T_{AB}^p T_{CD}^q.
\ee Here, $A,B,C,D$ stand for the indices of the representation $R_0$,
$p,q$ are the adjoint indices of gauge group $G$,
 $T^p_{AB}$ corresponding generators and
$K_{pq}$ the inverse of the Chern-Simons coefficient.
This superpotential vanishes if \begin{equation}
f_{A(BCD)}=0,\label{GW}
\end{equation} then one has the $U(1)_F$ flavor symmetry
which rotates the entire superfield $Q\to e^{i\theta} Q$,
i.e.~it acts as $(q,\psi_q)\to(e^{i\theta} q, e^{i\theta }\psi_q)$ where
$q$ and $\psi_q$ are the lowest and the fermion component of $Q$, respectively.
Now, the Cartan part of $SU(2)_R$ symmetry acts as
$(q,\psi_q)\to(e^{i\theta} q, e^{-i\theta }\psi_q)$.
Thus there are two $U(1)_R$ symmetries which
rotate the lowest component and the fermion component separately.
It means that the R-symmetry should be enhanced from $SO(3)_R$ to
$SO(4)_R$, and
the resulting theory has $\cN=4$ supersymmetry.
This condition is what was found by \cite{GW}.
The superpotential in the $\cN=2$ superfield formalism is zero, but this theory has
non-trivial sextic scalar coupling from integrating out of the auxiliary fields in the vector multiplet,
see \cite{YG}.

Now let us put two hypermultiplets $Q_{1,2}$ in the same pseudoreal representation $R_0$.
There is an $SO(2)_F$ flavor symmetry which rotates the two. The superpotential after the elimination
of the auxiliaries is \begin{equation}
W=f_{ABCD}(Q^A_1 Q^B_1+Q^A_2 Q^B_2)( Q^C_1 Q^D_1+ Q^C_2 Q^D_2),
\end{equation} which becomes under the assumption \eqref{GW} \begin{equation}
W=2f_{ABCD}Q^A_1 Q^B_1Q^C_2 Q^D_2
= 2 f_{ABCD}\epsilon^{\alpha\gamma}\epsilon^{\beta\delta}Q^A_\alpha Q^B_\beta Q^C_\gamma Q^D_\delta.
\end{equation} 
This makes manifest that there is an enhancement of the flavor symmetry from $SO(2)_F$ to $SU(2)_F$, 
which combine with the $SU(2)_R$ symmetry  to form $SO(5)_R$ symmetry
of the $\cN=5$ supersymmetry. This is also discussed by \cite{H2} in a different language.

Next, consider the case with chiral multiplets $A_i$, $B_i$ ($i=1,2$)
in the representation $R$ and $\bar R$, respectively.
It has a manifest  $SU(2)_F$ symmetry which acts on $A_i$ and $B_i$ as doublets.
Elimination of the auxiliary fields induces the superpotential \begin{equation}
W=f_{a\bar bc\bar d}(A_1^a B_1^{\bar b}+A_2^a B_2^{\bar b})(A_1^c B_1^{\bar d}+A_2^c B_2^{\bar d})
\end{equation}
where
\be
f_{a\bar bc\bar d} = K_{pq} T_{a\bar b}^p T_{c\bar d}^q.\label{definition-of-f}
\ee Here, $a,c$ stand for the indices of the representation $R$, $\bar b,\bar d$ those for $\bar R$,
$p,q$ are the adjoint indices of gauge group $G$, $T^p_{a\bar b}$ the corresponding generator on $R$,
and $K_{pq}$  the inverse Chern-Simons coefficient.
Now suppose $f_{a\bar b c\bar d}$ is antisymmetric in $a$ and $c$. Then the superpotential becomes
\begin{equation}
W=2f_{a\bar bc\bar d}A_1^a B_1^b A_2^a B_2^b
\end{equation} which shows the existence of the $SU(2)_A\times SU(2)_B$ symmetry
acting independently on $A_i$ and $B_i$.  This symmetry does not commute with the $SU(2)_R$
symmetry, because the latter rotates $A_1$ to $B_1^\dagger$.
Therefore they combine to  form the $SU(4)_R=SO(6)_R$ symmetry for $\cN=6$ theory.
This is the mechanism found by \cite{ABJM}.

Assume furthermore $R$ is a strictly real representation $R=\bar R$.
Then one can forget the distinction between indices $a$ and $\bar b$ and
$f_{abcd}$ becomes totally antisymmetric in four indices,
and there is the symmetry $SU(4)_F$
which rotates four fields $A_{1,2}$, $B_{1,2}$. In this case
the supersymmetry enhances to $\cN=8$.  This object $f_{abcd}$ is exactly the
structure constant of 3-algebra which automatically satisfies the celebrated fundamental identity.

In view of these facts our goal
is to classify all possible gauge groups and matter representation that give rise to this type of the enhancement
to $\cN=6$.
As a byproduct we obtain a totally independent uniqueness proof
 of $\cN=8$ theory with positive kinetic terms \cite{papa,GG}.

The rest of the paper is devoted to the classification. In
Sec.~\ref{classification} we find that the only allowed irreducible
representations allowed are either \begin{itemize}
\item Bifundamental of $SU(n)\times SU(n)$ ,
\item Bifundamental of $SU(m)\times SU(n)$ ($m\ne n$) with a particular $U(1)$
charge
\item Fundamental of $Sp(n)$ or $SU(n)$ with a particular $U(1)$ charge.
\end{itemize}
We also show that reducible representations do not offer any new possibilities.
We conclude the paper in Sec.~\ref{N8} with a comment on the classification of $\cN=8$ theories,
which follows by examining the list above to search for a strictly real representation.

{\bf Note added:} \  In the last stage of the preparation of the paper,
after having read the paper \cite{H2} appeared last week, the authors understood that the
classification of $\cN=6$ Lagrangian of the ABJM type can be reduced to the classification of
Lie superalgebras \cite{Kac}
whose fermionic part is a direct sum of a representation and its complex conjugate.
We believe our work offers a complementary  and useful perspective on the classification problem.

\section{Classification}
\label{classification}

The condition for supersymmetry enhancement for a general gauge group $G=G_1
\times G_2 \times \cdots \times G_L$ is that $f_{a\bar b c\bar d}$ , defined in
\eqref{definition-of-f}, be antisymmetric in the $a$ and $c$ indices. Of course
it should be gauge invariant as well, this demands that $K_{pq}$ be
proportional to the Cartan-Killing metric on each gauge group factor, possibly
with different coefficients. When we normalize the Chern-Simons term in the
action for each of the group factors as
\begin{equation}
S=\int \frac {k}{8\pi} g^{pq} A_p dA_q + \text{cubic terms},
\end{equation}
we have \begin{equation}
f_{a\bar bc\bar d} = \sum_{l=1}^L \frac{4\pi}{k_l} g_{pq}^{(l)}T_{(l) a\bar b}^p T_{(l) c\bar d}^p
\end{equation} as was determined in \cite{ABJM}
\footnote{
The quantization condition of $k_l$ depends on the normalization of $g_{pq}$
and also the discrete quotients. We identify the root space with the Cartan subalgebra, 
which has a natural inner product inherited from $g_{pq}$. 
Thus the normalization of the length of the roots determines the quantization of $k_l$. $k_l$ is integrally quantised for the simply-connected group
when the long roots have squared length two, which is the normalization we employ.}.
We take the orthonormal basis of the Lie algebra to be
\be
H_{(l)}^i,\quad \frac{|\alpha|}{2} (E_{(l)}^\alpha + E_{(l)}^{-\alpha}),\quad
\frac{|\alpha|}{2i} (E_{(l)}^\alpha - E_{(l)} ^{-\alpha}),
\ee
where $H^i$, $E^{+\alpha}$ and $E^{-\alpha}=(E^{+\alpha})^\dagger$
satisfy the commutation relation \footnote{We omit the subscript $l$ when there is
no confusion. Generators with different $l$ commute with each other. We follow the notation of \cite{CFT}.}
\bea
\left[H^i, H^j\right] &=& 0, \\
\left[H^i, E^\alpha\right] &=& \alpha^i E^\alpha, \\
\left[ E^\alpha, E^\beta\right] &=& N_{\alpha,\beta} E^{\alpha+\beta} \quad {\rm if}\, \alpha+\beta \in \Delta, \\
 &=& \frac{2}{|\alpha|^2} \alpha^i H^i  \quad {\rm if}\, \alpha+\beta =0, \\
 &=& 0 \quad {\rm otherwise} .
\eea
Here $\Delta$ is the set of roots.
We think of these generators as given in a particular representation and we
shall use `quantum mechanical' notation. Instead of writing $T_{a\bar b}$ we shall
write $\aver{b|T|a}$. This will allow us to study the antisymmetry property in
a basis independent and more powerful way.

A generic finite dimensional representation of a Lie algebra $A$ starts with a
highest weight vector which we shall call $\ket{\lambda}$. It has the property
that
\bea
H^{i} \ket{\lambda} &=& \lambda^i \ket{\lambda} \\
E^\alpha \ket{\lambda} &=& 0, \quad  \alpha \in \Delta_+,
\eea
where $\Delta_+$ denotes the set of positive roots. It can be of course
annihilated by some $E^{-\alpha}$ with negative roots, depending on a
particular representation. We shall normalize its inner product
$\aver{\lambda|\lambda}=1$. This product should not be confused with the
product on the root or weight spaces denoted by round brackets. In this
notation our object of interest is
\bea
f_{a\bar bc\bar d} &=& \sum_{l=1}^L \frac{4\pi}{k_l}g_{pq}^{(l)} \aver{b| T_{(l) }^p |a} \aver{d|
T_{(l)}^p|c} \\
&=&  \sum_{l=1}^L \frac{4\pi}{k_l} \left[ \sum_i \aver{b| H_{(l) }^i |a} \aver{d|
H_{(l)}^i|c} + \sum_{\alpha \in \Delta} \frac{|\alpha|^2}{2} \aver{b| E_{(l)
}^{\alpha} |a} \aver{d| E_{(l)}^{-\alpha}|c} \right].
\eea
Hereafter we redefine $f$ by a factor of $4\pi$.

Let us start by considering $f_{\lambda\bar \lambda\lambda\bar \lambda}$. By antisymmetry
this should be zero, but direct computation reveals
\be
f_{\lambda\bar \lambda\lambda\bar \lambda} =  \sum_{l=1}^L \frac{1}{k_l}
(\lambda,\lambda)_l,
\ee
this imposes a single constraint on the Chern-Simons levels $k_l$ of the
individual gauge groups, and allowed representations. In the case of single
semisimple group factor this will turn out to be the only constraint on the
levels coming from classical and not quantum considerations.

\begin{figure}
\centerline{\includegraphics[width=.2\textwidth]{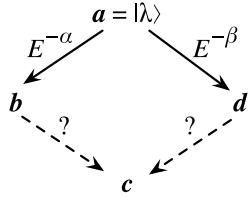}}
\caption{Choice of the states for \eqref{bra-eq}\label{states}}
\end{figure}

For the next step, let us take three states $\ket{a}=\ket{\lambda}$,
$\ket{b}=E^{-\alpha} \ket{\lambda}$ and $\ket{d}=E^{-\beta} \ket{\lambda}$, see
Fig.~\ref{states}. We shall leave $\ket{c}$ arbitrary. Demanding that $f_{a\bar
bc\bar d} + f_{c\bar ba\bar d}=0$ for all possible vectors $\ket{c}$ implies
\be\label{bra-eq}
(\alpha,\lambda) \bra{\lambda}E^\beta E^\alpha + (\beta,\lambda) \bra{\lambda}
 E^\alpha E^\beta =0.
\ee
In particular for $\alpha=\beta$ this implies either $(\alpha,\lambda) =0$ or
$\bra{\lambda} \left(E^\alpha\right)^2 =0$. The latter condition implies
\bea
0 &=& \aver{\lambda|\left(E^\alpha\right)^2\left(E^{-\alpha}\right)^2|\lambda}
\nonumber\\
&=& (\alpha,2\lambda-\alpha)(\alpha,\lambda) \frac{4}{|\alpha|^4},
\eea
which in turn implies
\be\label{DynkinConstr}
\frac{2(\alpha,\lambda)}{|\alpha|^2} = 0 \; {\rm or} \; 1.
\ee
The combination on the left hand side when evaluated on simple roots is the
definition of the Dynkin labels which characterize the Lie algebra
representations. The equation (\ref{DynkinConstr}) was not derived for simple
roots only, but for all positive roots. Let $\alpha_i$ and $\alpha_j$ be two
simple roots adjacent on the Dynkin diagram. The sum of two adjacent simple
roots is always a root.\footnote{Given two roots $\alpha$ and $\beta$ their sum
is always a root, as long as $(\alpha,\beta)$ is negative. This condition is
always satisfied for adjacent simple roots. This and other properties of roots
can be found in \cite{FH}.} Let us then take $\alpha =
\alpha_i+\alpha_j$ and examine (\ref{DynkinConstr}). We will now show by
contradiction, that at most one of the two adjacent Dynkin labels is nonzero.
Suppose then, that both labels are equal one. The factor
$(\alpha_i+\alpha_j,\lambda)$ equals $(|\alpha_i|^2+|\alpha_j|^2)/2 >0$ and the
the second factor $(\alpha,2\lambda-\alpha)$  becomes
$|\alpha_i|^2+|\alpha_j|^2 -|\alpha_i+\alpha_j|^2$. This can only vanish when
$(\alpha_i,\alpha_j)=0$ which is never true for adjacent simple roots.

Let us now come back to (\ref{bra-eq}) and take $\alpha$ and $\beta$ to be two
different, nonadjacent simple roots. Their sum is never a root, the
corresponding operators $E^\alpha$ and $E^\beta$ thus commute and we get a
condition $(\alpha+\beta,\lambda)=0$ or $\bra{\lambda}  E^\alpha E^\beta =0$.
The former condition can be satisfied only when $(\alpha,\lambda)=0$ and
$(\beta,\lambda)=0$ hold simultaneously. The latter condition is weaker and is
equivalent to $(\alpha,\lambda)=0$ or $(\beta,\lambda)=0$. This shows that
given a pair of two nonadjacent simple roots, at most one carries a nonzero
Dynkin label. Combining this with our result from the previous paragraph this
shows that there is exactly single nonzero label in the whole Dynkin diagram,
which furthermore has to be equal to one. Such representations are called
minuscule representations, but they will not play a role here. In the next
paragraph we will show that the only allowed place for the unit label is one of
the terminal nodes of the Dynkin diagram.

\subsection{Only Dynkin labels at the terminal  node are allowed}

\begin{figure}
\centerline{
\includegraphics[width=.3\textwidth]{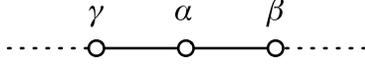}
}
\caption{Three adjacent roots\label{adjacent-dynkin}}
\end{figure}

\begin{figure}
\centerline{
\includegraphics[width=.2\textwidth]{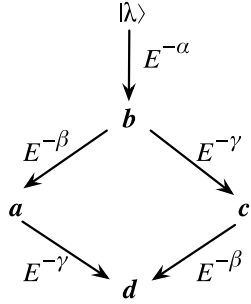}
}
\caption{Choice of states for \eqref{eq_terminal}\label{adjacent-states}}
\end{figure}

Let us take three adjacent simple roots $\alpha, \beta, \gamma$, with $\alpha$
being the middle one, i.e. $(\alpha,\beta) < 0$, $(\alpha,\gamma) < 0$ and
$(\beta,\gamma) = 0$, see Fig.~\ref{adjacent-dynkin}.
 Let us assume the Dynkin label of the middle root being
$1$, i.e. $(\lambda,\alpha)=|\alpha|^2/2$ , whereas the Dynkin labels of
$\beta$ and $\gamma$ are zero, that is $(\lambda,\beta)=(\lambda,\gamma)=0$.

Consider four nontrivial states in this representation given by $\ket{a} =
E^{-(\alpha+\beta)} \ket{\lambda}, \; \ket{b} = E^{-\alpha} \ket{\lambda}, \;
\ket{c} = E^{-(\alpha+\gamma)} \ket{\lambda},\; \ket{d} =
E^{-(\alpha+\beta+\gamma)} \ket{\lambda} $, also see Fig.~\ref{adjacent-states}.
 Simple computation shows that
\be\label{eq_terminal}
f_{a\bar bc\bar d}+f_{c\bar ba\bar d}= \frac{2}{k}
\frac{(\lambda,\alpha)^2}{|\alpha+\beta+\gamma|^2}
\left(N_{\alpha,-(\alpha+\beta)} N_{-\beta,-(\alpha+\gamma)} +
N_{\alpha,-(\alpha+\gamma)} N_{-\gamma,-(\alpha+\beta)}\right),
\ee
where we used an identity
\be
N_{\alpha,\beta} = \frac{|\alpha+\beta|^2}{|\beta|^2}
N_{\alpha,-(\alpha+\beta)} = \frac{|\alpha+\beta|^2}{|\alpha|^2}
N_{\beta,-(\alpha+\beta)}
\ee
valid for any two roots $\alpha$ and $\beta$ whose sum is a root as well. From
the Jacobi identity
\be
\left[\left[E^\alpha,E^{-(\alpha+\beta)}\right],E^{-(\alpha+\gamma)}\right] =
\left[\left[E^\alpha,E^{-(\alpha+\gamma)}\right],E^{-(\alpha+\beta)}\right] +
\left[\left[E^{-(\alpha+\gamma)},E^{-(\alpha+\beta)}\right],E^\alpha\right]
\ee
we learn that
\be
N_{\alpha,-(\alpha+\beta)} N_{-\beta,-(\alpha+\gamma)} =
N_{\alpha,-(\alpha+\gamma)} N_{-\gamma,-(\alpha+\beta)}
+N_{-(\alpha+\gamma),-(\alpha+\beta)} N_{-(2\alpha+\beta+\gamma),\alpha},
\ee
where the last term vanishes or rather is not present if $2\alpha+\beta+\gamma$
is not a root. In such a case the two terms in (\ref{eq_terminal}) are equal to
each other, and from the general theory of Lie algebras, they are nonzero. What
are the actual sequences of adjacent simple roots $(\beta,\alpha,\gamma)$, such
that $2\alpha+\beta+\gamma$ is a root? From the four possible orderings of the
long and short roots (up to reversal symmetry) $(L,L,L), (S,S,S), (S,L,L),
(S,S,L)$, only the last sequence gives rise to $2\alpha+\beta+\gamma$ being a
root. This can be established by computing the scalar product
$(\alpha+\beta,\alpha+\gamma)$. For the first three sequences the product is
zero, and since the difference $(\alpha+\beta)-(\alpha+\gamma)$ is not a root,
neither is the sum. In the $(S,S,L)$ case the scalar product equals $-1/2$ and
so the sum is a root. This rules out antisymmetry of $f_{a\bar bc\bar d}$ in
cases when the nonzero Dynkin label is the middle of a sequence of three simple
roots $(L,L,L), (S,S,S), (S,L,L)$.

What about the $(S,S,L)$ case? To exclude possible cancelation of the two terms
in (\ref{eq_terminal}) we compute absolute value squared of each of them using
a formula
\be\label{N_id}
\frac{|N_{\alpha,\beta}|^2}{|\alpha+\beta|^2} -
\frac{|N_{\alpha,-\beta}|^2}{|\alpha-\beta|^2} =-
2\frac{(\alpha,\beta)}{|\alpha|^2|\beta|^2}
\ee
that can be easily obtained from the Jacobi identities for $E^\alpha,
E^{-\alpha}, E^{-\beta}$ and $H^i, E^\alpha, E^{-(\alpha+\beta)}$, and using
that $N_{\alpha,\beta}^* = -N_{-\alpha,-\beta}$ in our basis wherein
$\left(E^{\alpha}\right)^\dagger = E^{-\alpha}$. From this one can derive for
the adjacent simple roots
\bea
|N_{\alpha,-(\alpha+\beta)}|^2 &=& - \frac{2(\alpha,\beta)}{|\alpha+\beta|^2} \frac{|\beta|^2}{|\alpha|^2} \\
|N_{-\beta,-(\alpha+\gamma)}|^2 &=& -
2\frac{(\beta,\alpha+\gamma)}{|\beta|^2|\alpha+\gamma|^2}
|\alpha+\beta+\gamma|^2
\eea
by using the fact that $\alpha-\beta$ and $\beta-(\alpha+\gamma)$ can never be
roots if $\alpha, \beta$ and $\gamma$ are simple.
 To show that the two terms in (\ref{eq_terminal}) cannot cancel in the
 $(S,S,L)$ case, it suffices to prove that
 \be
 (\alpha,\beta)(\beta,\alpha+\gamma) \ne (\alpha,\gamma)(\gamma,\alpha+\beta),
\ee
which is manifestly true since $(\beta,\gamma)=0$ and $(\alpha,\beta) =
(\alpha,\gamma)/2$. We have thus succeeded showing that the only nonzero Dynkin
label (being necessarily a unity), must sit at one of the ends of a Dynkin
diagram. In the next section we will study what kind of Dynkin diagrams are
allowed.

\subsection{No branches in the Dynkin diagram allowed}

\label{no-branches}

Let us star by recalling a basic fact from the Cartan-Weyl theory. Given any
sequence of adjacent simple roots $\alpha_1, \alpha_2, \ldots, \alpha_n$, their
sum is again a root. It is easiest to see this by induction:
\be
\sum_{i=1}^{k-1} \left( \alpha_i , \alpha_k \right) = (\alpha_{k-1},\alpha_k)
<0,
\ee
the left hand side being $-1/2$ or $-1$. Therefore their sum is a root as well.

\begin{figure}
\centerline{
\includegraphics[width=.3\textwidth]{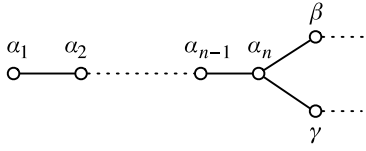}
}
\caption{Branched Dynkin diagram\label{branch-dynkin}}
\end{figure}

Now consider a $D$ or $E$ type Dynkin diagram with a sequence of simple roots
$\alpha_1, \alpha_2, \ldots, \alpha_n$, where $\alpha_1$ is a root at one of
the ends of the diagram, and let us further assume, that there are two
additional simple roots $\beta$ and $\gamma$ that are adjacent to $\alpha_n$
(apart of $\alpha_{n-1}$ of course). See Fig.~\ref{branch-dynkin}.
 Let us consider a root
\be
\alpha=\sum_{k=1}^{n} \alpha_k.
\ee
In all the simply laced Dynkin diagram (the only ones that can have branches)
all roots, not necessarily simple ones, have length squared $2$. From what we
have said, $\alpha$, $\alpha+\beta$, $\alpha+\gamma$, $\alpha+\beta+\gamma$ are
all roots. On the other hand $2\alpha+\beta+\gamma$ is not a root. This is
because the scalar product $(\alpha,\alpha+\beta+\gamma)=2-1-1=0$ and hence
$2\alpha+\beta+\gamma$ could be a root if and only if
$\alpha+\beta+\gamma-\alpha=\beta+\gamma$ was a root, which it is not. Sum of
two orthogonal simple roots is never a root.

 Just as in the previous subsection we can take $\ket{a} = E^{-(\alpha+\beta)}
\ket{\lambda}, \; \ket{b} = E^{-\alpha} \ket{\lambda}, \; \ket{c} =
E^{-(\alpha+\gamma)} \ket{\lambda},\; \ket{d} = E^{-(\alpha+\beta+\gamma)}
\ket{\lambda} $ and evaluate $f_{a\bar bc\bar d}+f_{c\bar ba\bar d}$, see Fig.~\ref{adjacent-states}.
 Actually we do not need to to
do any new computation, since all that was used to derive (\ref{eq_terminal})
was that $(\lambda,\beta)=(\lambda,\gamma)=0$. As we showed in the previous
paragraph, $2\alpha+\beta+\gamma$ is not a root, and therefore as explained in
the previous section by the use of the Jacobi identity the two terms
(\ref{eq_terminal}) are exactly the same including the sign, and each of them
non-zero. This concludes the proof for the $D$ and $E$ type diagrams, that they
cannot lead to $f_{a\bar bc\bar d}$ antisymmetric in $a$ and $c$.

\subsection{No short roots allowed if a Dynkin label is on a long root}
\label{short}
We are now going to show that $B_n$, $C_n$, $G_2$ and $F_4$ diagrams cannot
have a nonzero Dynkin label on the long-root end. Let $\alpha$ be a root given
by a sum of all the simple roots of length squared 2. (Such roots form
necessarily a connected set of nodes with no branches, and therefore their sum
is again a root.) Let $\beta$ be the first short root. We have $|\alpha|^2=2$,
$|\beta|^2=1$ (or $2/3$ in case of $G_2$), $(\alpha,\beta)=-1$. A crucial fact
is that $\alpha+2\beta$ is another root since $(\alpha+\beta,\beta) \le 0$  and
the difference is a root, hence so must be the sum.

\begin{figure}
\centerline{
\includegraphics[width=.2\textwidth]{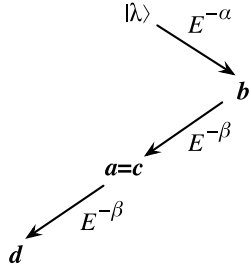}
}
\caption{Choice of states for Sec.~\ref{short}\label{short-states}}
\end{figure}

Assigning Dynkin label $1$ to the long root at one of the ends implies
$(\alpha,\lambda)=1$, $(\beta,\lambda)=0$. The states $\ket{a} =
E^{-(\alpha+\beta)} \ket{\lambda}, \; \ket{b} = E^{-\alpha} \ket{\lambda}, \;
\ket{c} = E^{-(\alpha+\beta)} \ket{\lambda},\; \ket{d} = E^{-(\alpha+2\beta)}
\ket{\lambda} $ are all nontrivial, see Fig.~\ref{short-states}.
 To compute the $f_{a\bar bc\bar d}$ we do not need to
do any new computation, we can again use formula (\ref{eq_terminal}) and set
$\gamma=\beta$. The $f_{a\bar ba\bar d}$ is manifestly nonzero which violates the
antisymmetry.

\subsection{At most single long root if a Dynkin label on a short root}
\label{short2} We are now going to prove that a Dynkin diagram having a unit
label on one of its terminal nodes representing a short simple root, cannot
have more than one long simple root. This will eliminate the spin
representation of $SO(2n+1)$ for $n\ge3$ and the $26$ of $F_4$. Presenting a
separate argument for the {\bf 7}=$(0,1)$ of $G_2$ later on for completeness
will leave us with (anti)fundamental representation of the $A_n$ algebras and
the ${\bf 2n}=(1,0,...0,0)$ of $C_n=Sp(n)$.

\begin{figure}
\centerline{
\includegraphics[width=.2\textwidth]{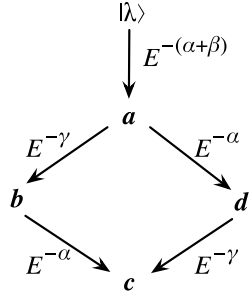}
}
\caption{Choice of states for Sec.~\ref{short2}\label{short-states-2}}
\end{figure}

Let $\alpha$ denote the sum of all the short roots in the Dynkin diagram. Let
$\beta$ be the first long root at the end of the short-root sequence, and let
$\gamma$ be another long root adjacent to $\beta$. We thus have $|\alpha|^2=1$,
$|\beta|^2=|\gamma|^2=2$ and $(\alpha,\beta)=(\beta,\gamma)=-1$ and
$(\alpha,\gamma)=0$. By the by now familiar argument $\alpha+\beta,
2\alpha+\beta, \alpha+\beta+\gamma$ and $2\alpha+\beta+\gamma$ are all positive
roots. Since the Dynkin label sits on one of the simple roots within $\alpha$,
it follows that $(\lambda,\alpha)=1/2$ whereas
$(\lambda,\beta)=(\lambda,\gamma)=0$. The states $\ket{a} = E^{-(\alpha+\beta)}
\ket{\lambda}, \; \ket{b} = E^{-(\alpha+\beta+\gamma)} \ket{\lambda}, \;
\ket{c} = E^{-(2\alpha+\beta+\gamma)} \ket{\lambda},\; \ket{d} =
E^{-(2\alpha+\beta)} \ket{\lambda} $ are thus all nontrivial, see Fig.~\ref{short-states-2}.
 Computing
$f_{a\bar bc\bar d}$ we find
\bea
f_{a\bar bc\bar d} &=&\frac{2}{k} \frac{|\gamma|^2
(\alpha+\beta+\gamma,\lambda)(2\alpha+\beta+\gamma,\lambda)}{|\alpha+\beta+\gamma|^2|2\alpha+\beta+\gamma|^2}
N_{-\gamma,-(\alpha+\beta)} N_{2\alpha+\beta,\gamma}\\
&=&\frac{1}{k} N_{-\gamma,-(\alpha+\beta)} N_{2\alpha+\beta,\gamma}
\eea
and from (\ref{N_id}) we find the absolute value $|f_{a\bar bc\bar d}|=1/|k|$.
For $f_{c\bar ba\bar d}$ we get
\bea
f_{c\bar ba\bar d} &=&\frac{2}{k} \frac{|\alpha|^2
(\alpha+\beta+\gamma,\lambda)(\alpha+\beta,\lambda)}{|\alpha+\beta+\gamma|^2|\alpha+\beta|^2}
N_{\alpha,-(2\alpha+\beta+\gamma)} N_{2\alpha+\beta,-\alpha}\\
&=&\frac{1}{2k} N_{\alpha,-(2\alpha+\beta+\gamma)} N_{2\alpha+\beta,-\alpha}
\eea
and hence $|f_{c\bar ba\bar d}|=1/2|k|$. So this violates antisymmetry.

\begin{figure}
\centerline{\includegraphics[width=.3\textwidth]{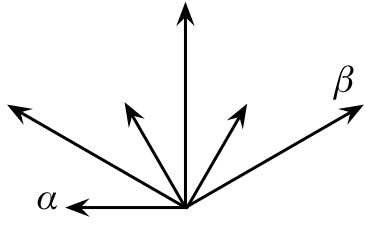}}
\caption{Positive roots of $G_2$.\label{g2-roots}}
\end{figure}

\begin{figure}
\centerline{\includegraphics[width=.3\textwidth]{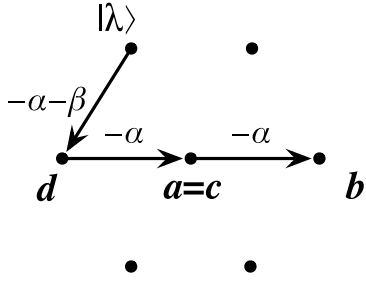}}
\caption{Choice of states for $\bf 7$ of $G_2$\label{g2-states}}
\end{figure}

To finish this discussion we show that $G_2$ is not a good candidate either.
$G_2$ has six positive roots, write them as $\alpha, \beta, \alpha+\beta,
2\alpha+\beta, 3\alpha+\beta$ and $3\alpha+2\beta$ of lengths squared $2/3, 2,
2/3,2/3,2$ and $2$ respectively; see Fig.~\ref{g2-roots}.
Take the following states: $\ket{a} =\ket{c} =
E^{-(2\alpha+\beta)} \ket{\lambda}, \; \ket{b} = E^{-(3\alpha+\beta)}
\ket{\lambda}, \; \ket{d} = E^{-(\alpha+\beta)} \ket{\lambda} $ which are
nontrivial in the ${\bf 7}=(0,1)$ representation, with $\alpha$ denoting the
short simple root, see Fig.~\ref{g2-states}.
 The $f_{a\bar ba\bar d}$ is readily computed, it is given by
\be
f_{a\bar ba\bar d}=\frac{1}{k} \frac{|\alpha|^2}{2} \aver{\lambda|E^{3\alpha+\beta}
E^{-\alpha} E^{-(2\alpha+\beta)}|\lambda} \aver{\lambda|E^{\alpha+\beta}
E^{\alpha} E^{-(2\alpha+\beta)}|\lambda}
\ee
which by now familiar arguments is obviously nonzero and hence $G_2$ is not a
good candidate either. To conclude, the only groups that appear admissible are
the $A_n=SU(n+1)$ in ${\bf n+1}$ or ${\bf\overline{n+1}}$, and $C_n=Sp(n)$ in
the ${\bf 2n}$ representation. To prove that they can appear we now have to
understand what products of different groups are allowed.

\subsection{Patching several gauge groups}

Let us look into constraints arising when having multiple gauge groups. We can
slightly modify the computation that gave rise to equation (\ref{bra-eq}) by
considering the states $\ket{a}=\ket{\lambda}$, $\ket{b}=E_{l_1}^{-\alpha}
\ket{\lambda}$ and $\ket{d}=E_{l_2}^{-\beta} \ket{\lambda}$, where now $l_1$
and $l_2$ label two distinct simple gauge groups. Demanding antisymmetry of
$f_{a\bar bc\bar d}$ we find the analog of (\ref{bra-eq}) to be
\be\label{bra-eq2}
\left(\frac{1}{k_{l_1}} (\alpha,\lambda)_{l_1} +
\frac{1}{k_{l_2}}(\beta,\lambda)_{l_2} \right) \bra{\lambda}
E_{l_1}^{\alpha}E_{l_2}^{\beta} =0.
\ee
If the representation $R$ is represented nontrivially on both gauge groups, and
one takes the roots $\alpha$ and $\beta$ to have nonzero overlap with the
highest weight $\lambda$ then the state $ E_{l_1}^{-\alpha}E_{l_2}^{-\beta}
\ket{\lambda}$ is nontrivial. One could take for instance the simple roots that
specify the Dynkin labels. Fixing the root $\beta$, such that
$(\beta,\lambda)_{l_2} \ne 0$, one can now vary the root $\alpha$ over all
roots that satisfy $(\alpha,\lambda)_{l_1} \ne 0$ and from (\ref{bra-eq2}) we
see that $(\alpha,\lambda)$ must be constant for all the roots! This is not a
strong requirement for the $A_n=SU(n+1)$ groups, where all the roots have
length squared $2$, but for the $C_n=Sp(n)$ groups it is. We have already seen
that in the fundamental representation there are roots having a nonvanishing
overlap with the highest weight of both lengths, long and short. Let $\alpha$
be the sum of all short simple roots and $\alpha_n$ be the only simple long
root. This root is orthogonal to $\lambda$, but it can be used to construct
another root $2\alpha+\alpha_n$ of length squared $2$ that is not orthogonal to
$\lambda$, but instead $(2\alpha+\alpha_n,\lambda)=1$ whereas
$(\alpha,\lambda)=1/2$. This shows that in a given irreducible component of the
representation there can be no field that transforms nontrivially under $Sp(n)$
and another semisimple gauge group.

Let us now suppose there were three or more simple gauge groups represented
nontrivially in a single irreducible representation. Then, provided
$(\alpha,\lambda)_{l_1} \ne 0, (\beta,\lambda)_{l_2} \ne 0$ and $
(\gamma,\lambda)_{l_3} \ne 0$ we would find
\bea
\frac{1}{k_{l_1}} (\alpha,\lambda)_{l_1} +
\frac{1}{k_{l_2}}(\beta,\lambda)_{l_2}  &=& 0,
\nonumber \\
\frac{1}{k_{l_2}} (\beta,\lambda)_{l_2} +
\frac{1}{k_{l_3}}(\gamma,\lambda)_{l_3}  &=& 0,
\nonumber \\
\frac{1}{k_{l_1}} (\alpha,\lambda)_{l_1} +
\frac{1}{k_{l_3}}(\gamma,\lambda)_{l_3}  &=& 0,
\eea
which is a contradiction.

To conclude we have found that any given field transforms nontrivially at most
under two simple groups in case of $SU(n)$ or under a single $Sp(n)$, up to
additional abelian $U(1)$ factors that might be needed to make $\sum_l
k_l^{-1}(\lambda,\lambda)$ vanish. In case of $SU(n)\times SU(m)$ we have
$(\alpha,\lambda)=(\beta,\lambda)=1$ and the corresponding levels must be
opposite to each other. The condition $\sum_l k_l^{-1}(\lambda,\lambda)=0$
would not however be satisfied if $n \ne m$, since once the length-squared of
the long roots is set to two, the length-squared of the fundamental weight for
$SU(n)$ is equal to $(n-1)/n$ and would be different for both factors. The
minimal gauge groups that are allowed are $SU(n) \times U(1)$, $Sp(n) \times
U(1)$, $SU(n) \times SU(n)$ and $SU(n) \times SU(m) \times U(1)$. Additional
$U(1)$'s can be added, but no other semi-simple factors.

\subsection{Adding $U(1)$ factors}

In the presence of several $U(1)$'s, it might not be desirable to diagonalize
the part of the matrix $K_{pq}$ appearing in front of the abelian generators in
(\ref{f-def}), since this could introduce non-integer values for the $U(1)$
charges. Let us write our gauge group as $G=G_\text{semi-simple} \times U(1)_1
\times \ldots \times U(1)_{M}$. The tensor $f_{a\bar bc\bar d}$ takes up a form
\be\label{f-decomp}
f_{a\bar bc\bar d} = f_{a\bar bc\bar d}^\text{semi-simple} + \delta_{a\bar b}\delta_{c\bar d}\, K_{mn} q^m q^n,
\ee
where $q^m$ are charges of the representation under the $m$-th $U(1)$. From the
point of view of antisymmetry of $f_{a\bar bc\bar d}$ in $a$ and $c$, it is
immaterial how many $U(1)$'s there are. The only important thing is that the
constant $K_{mn} q^m q^n$ has the appropriate value to cancel a
non-antisymmetric part of $f_{a\bar bc\bar d}^\text{semi-simple}$. Contracting
$f_{a\bar bc\bar d}+f_{c\bar ba\bar d}$ with $\delta_{a\bar b} \delta_{c\bar
d}$ and using the normalization for the trace in the fundamental representation
\be
\Tr_{\lambda} T^p T^q = \delta^{pq}
\ee
we find
\be
K_{mn} q^m q^n = -\frac{1}{\dim R(\dim R +1)} \sum_{l; G_l \, \text{simple}}
\frac{1}{k_l} \dim G_l.
\ee
Note that the normalization of the generators is fixed once we normalize the
length squared of the long roots to $2$. In a given representation $R_l$ of a
simple $G_l$ factor, given by the highest weight $\lambda$, the trace
normalization is
\be
\Tr_{\lambda} T_{(l)}^p T_{(l)}^q = \frac{\dim R_l
(\lambda,\lambda+2\rho)_l}{\dim G_l} \delta^{pq}.
\ee
Here $\rho$ is the Weyl vector given by the sum of all the fundamental weights
$\rho=\sum \omega_i$ which are defined to obey $(\omega_i,\alpha_j) =
\delta_{ij} |\alpha_j|^2/2$. In the fundamental representation the coefficient
on the right hand side equals 1.

In case the semi-simple part is actually simple, i.e. a single $SU(n)$ or
$Sp(n)$ it has to have a value of
\bea
K_{mn} q^m q^n &=& -\frac{1}{k} \frac{n-1}{n}, \qquad SU(n),
\\\label{K-Sp}
&=& -\frac{1}{2k} , \qquad\qquad Sp(n).
\eea

\subsection{Reducible representations}

So far we have analyzed irreducible representations only. Now we would like to
ask whether reducible representations are allowed, and if, what are the
conditions they must obey.

Let us start by assuming that a gauge group of the form $G=G_\text{semi-simple}
\times U(1)_1 \times \ldots \times U(1)_{M}$ has a finite dimensional reducible
representation $R=R_1 \oplus R_2 \oplus \ldots$. Let $A,B,C,D$ denote the
indices of $R_1$ and $\tilde A, \tilde B,\tilde C,\tilde D$ of the $R_2$
representation. The generators of the representation $R$ are block diagonal,
with $R_1$ and $R_2$ in the first two blocks (there might be other blocks as
well, but they are not important for our analysis). There is actually a very
strong constraint on possible representations. Consider
\be\label{fblock}
f_{A\bar B\tilde C\bartilde D} = \left(\sum_{l; G_l\, \text{simple}} \frac{1}{k_l} \bra{B}
T_{(l)}^p \ket{A} \bra{\tilde D} {\tilde T}_{(l)}^p \ket{\tilde C} \right) +
\delta_{A\bar B}\delta_{\tilde C \bartilde D}\, K_{mn} q^m {\tilde q}^n,
\ee
where $T$ and $\tilde T$ denote generators in the representation $R_1$ and
$R_2$ respectively; $q^m$ and $\tilde q^n$ are the $U(1)$ charges of both
representations. Because the generators are block diagonal and $\ket{B}$ and
$\ket{\tilde C}$ are orthogonal to each other, $f_{\tilde C \bar B A \bartilde D}$
automatically vanishes and therefore by antisymmetry $f_{A\bar B\tilde C\bartilde D}$
must be zero as well!

We are now going to prove that on every semisimple factor one of the two
representations must act trivially. Suppose on the contrary that there is a
generator of $G_l$ which is represented by nonzero matrices $T^p$ and $\tilde
T^p$ in the two representations. Let us take two arbitrary vectors $\ket{\tilde
C} \ne \ket{\tilde D}$ such that $\bra{\tilde D} {\tilde T}_{(l)}^p \ket{\tilde
C}$ is nonzero for at least a single generator ${\tilde T}_{(l)}^p$. This is
always possible, because non-abelian algebras cannot be represented by diagonal
matrices. The condition that (\ref{fblock}) vanishes can now be recast as
\be
\bra{B} \sum w_{(l)}^p T_{(l)}^p \ket{A} =0
\ee
for all vectors $\ket{A}$ and $\ket{B}$, and hence the set of generators of
$\left\{ T_{(l)}^p \right\}$ is linearly dependent which is a contradiction!

We have seen that if there are two representations of the gauge group $G$ they
cannot be simultaneously nontrivial on any simple factor. But can they be
non-trivial on a common abelian factor? The only condition arising from summing
two such irreducible representations is that the only surviving term in
(\ref{fblock}) vanishes as well,
\be\label{decoupling}
K_{mn} q^m {\tilde q}^n = 0.
\ee
It thus seems as if a theory based on a gauge group i.e. $SU(n)\times SU(m)
\times SU(n') \times SU(m') \times U(1)\times U(1)$ was allowed. The generators
would be in a reducible representation, in the first one, only the first two
simple gauge groups would be represented nontrivially, in the second
representation, the third and fourth group would be represented nontrivially.
One would have to merely solve a system of equations
\bea
\frac{1}{k_1}\left(\frac{n-1}{n} - \frac{m-1}{m}\right) + K_{ij}q^i q^j &=& 0,
\nonumber\\
\frac{1}{k_2}\left(\frac{n'-1}{n'} - \frac{m'-1}{m'}\right) + K_{ij}{\tilde
q}^i {\tilde q}^j &=& 0,
\nonumber\\
 K_{ij}{\tilde q}^i q^j &=& 0,
\eea
which surely is possible with enough $U(1)$'s (even though we have not
investigated whether solutions in integers exist). Upon a closer inspection
however, it turns out that the condition (\ref{decoupling}) enforces complete
decoupling between the two sectors of the theory. The matter and non-abelian
gauge sectors for the two representations are completely decoupled and the only
possible interaction coming from the shared abelian part effectively
disappears. Trying to write down any Feynman diagram connecting the two sectors
would involve a propagator of the abelian gauge fields containing the inverse
of the level matrix which is exactly our $K_{mn}$. This propagator would appear
between two currents of the two sectors and would couple with charges $q^m$ and
$\tilde q^n$ respectively. By the condition (\ref{decoupling})  the coupling
exactly vanishes.

\subsection{Existence}
Let us now check that the only possibilities $SU(n) \times U(1)$, $Sp(n) \times
U(1)$, $SU(n) \times SU(n)$ and $SU(n) \times SU(m) \times U(1)$ we found in
the previous sections indeed satisfy the fundamental identity.

\subsubsection{$SU(m)\times SU(n)\times U(1)$}
Here we treat the case of a bifundamental of $SU(m)\times SU(n)$, or a
fundamental of $SU(m)$.  The latter is a special case of the former with $n=1$.
We have $mn$ fields which are indexed by a pair $(a,\tilde a)$, where
$a=1,\ldots,m$  and $\tilde a=1,\ldots,n$. The generators of $SU(m)$ and
$SU(n)$ are \begin{equation} T^{p}_{a\tilde a,\bar b\bartilde b}=T^{p}_{a\bar
b}\delta_{\tilde a\bartilde b},\qquad T^{\tilde p}_{a\tilde a,\bar b\bartilde
b}=\delta_{a\bar b}T^{\tilde p}_{\tilde a\bartilde b}
\end{equation} for $p=1,\ldots,m^2-1$ and $\tilde p=1,\ldots,n^2-1$.
Here $T^p_{a\bar b}$ and $T^{\tilde p}_{\tilde a\bartilde b}$ are the
representation matrices in the fundamental representations. From the preceding
discussion we know that the levels of $SU(m)$ and $SU(n)$ are opposite to each
other. We denote the level of $SU(m)$ by $k$. Possible other $U(1)$ factors
acts by a scalar multiplication by $q_i$, with levels $k_i$. Recalling the fact
\begin{equation} \sum_{p=1}^{m^2-1} T^{p}_{a\bar b}T^{p}_{c\bar d} =
\delta_{a\bar d}\delta_{b\bar c}-\frac1m \delta_{a\bar b}\delta_{c\bar d},
\end{equation} we have
\bea\label{su-solution}
f_{a\tilde a,\bar b\bartilde b,c\tilde c,\bar d\bartilde d} &=& \frac1k\left(
    \delta_{a\bar d}\delta_{b\bar c}-\frac1m \delta_{a\bar b}\delta_{c\bar d}
\right) \delta_{\tilde a \bartilde b}\delta_{\tilde c\bartilde d} -\frac1k \delta_{ a
\bar b}\delta_{c \bar d} \left(
    \delta_{\tilde a\bartilde d}\delta_{\tilde c\bartilde b}
        -\frac1n \delta_{\tilde a\bartilde b}\delta_{\tilde c\bartilde d}
\right) + \\
    && +
(K_{ij}q^i q^j)
    \delta_{a\bar b}\delta_{c\bar d}\delta_{\tilde a\bartilde b}\delta_{\tilde c\bartilde d}
\eea
which is antisymmetric under the exchange $(a\tilde a)\leftrightarrow
(c\tilde c)$ if and only if \begin{equation}
K_{ij}q^i q^j=\frac1k\left(\frac1m-\frac1n\right).
\end{equation} For example it is satisfied if there are two $U(1)$ factors and
$q_{1,2}=1$, $K=\text{diag}(\frac{1}{km},\frac{1}{kn})$, which is exactly the case if we take a
bifundamental of $U(m)\times U(n)$ theory. The $\cN=6$ supersymmetry for
$U(m)\times U(n)$ with $m\ne n$
 was implicitly discussed in \cite{ABJM}; there, the discussion used the matrix notation for the bifundamentals $A_{1,2}$ and $B_{1,2}$ and they only mentioned the case $m=n$ explicitly.
But their derivation applies verbatim for $m\ne n$.

\subsubsection{$Sp(n)\times U(1)$}
Let us treat the case where we have $\mathbf{2n}$ of  $Sp(n)$. Let us denote
the generators of $Sp(n)$ by $T^p_{ab}$ where $a,b=1,\ldots,2n$. Extra $U(1)$
factors act by scalar multiplications by $q_i$, i.e.~ $iq_i \delta_a^b$ as matrices.
To lower the indices we use the standard invariant tensor $J_{ab}$ of $Sp(n)$,
therefore the generator $T^i_{ab}$ for the $i$-th $U(1)$ is given by
\begin{equation}
T^i_{ab}=i q_i  J_{ab}.
\end{equation}
We have
\begin{equation} \sum_{p=1}^{n(2n+1)} T^p_{ab}
T^p_{cd}=-\frac12(J_{ac}J_{bd}+J_{ad}J_{bc} ).
\end{equation} This relation follows, up to the constant factor, from
the symmetry consideration. Indeed, any invariant tensor of $Sp(n)$ can be
constructed from the product of $J$'s. Then we need to impose the symmetry in
$ab$ and $cd$, which uniquely fixes the index structure. Finally the
proportionality factor can be fixed by taking some component. Therefore we have
\begin{equation}\label{sp-solution}
 f_{abcd}=-\frac1{2k}(J_{ac}J_{bd}+J_{ad}J_{bc} ) -
(K_{ij}q^i q^j) J_{ab}J_{cd},
\end{equation} which is antisymmetric if and only if \begin{equation}
K_{ij}q^i q^j=-\frac1{2k}
\end{equation}
in full agreement with (\ref{K-Sp}).

\section{Comment on the uniqueness of the ${\cal N}=8$}
\label{N8}

The ABJM theory \cite{ABJM} discussed in this paper is expected to further
enhance its supersymmetry to ${\cal N}=8$  for certain values of the levels and
perhaps only for certain gauge groups. Such an enhancement has not been so far
exhibit manifestly and its detailed understanding presents an interesting
challenge.

Curiously, there is a unique theory constructed by Bagger, Lambert and
Gustavsson \cite{BL1,BL2,G} with manifest ${\cal N}=8$ supersymmetry, already
at the Lagrangian level. A class of such theories was written in terms of a
three-algebra $\left[T^a,T^b,T^c\right] = f_{\phantom{abc}d}^{abc} T^d$
required for consistency to satisfy rather stringent fundamental identity
\be\label{BLfund}
f_{\phantom{efg}d}^{efg}f_{\phantom{abc}g}^{abc}=f_{\phantom{efa}g}^{efa}f_{\phantom{bcg}d}^{bcg}
+f_{\phantom{efb}g}^{efb}f_{\phantom{cag}d}^{cag}+f_{\phantom{efc}g}^{efc}f_{\phantom{abg}d}^{abg}
\ee
in addition to complete antisymmetry in all the indices. It was shown in
\cite{papa,GG} that there is a single irreducible three algebra based on the
four-dimensional epsilon tensor that satisfies the fundamental identity. Here
we want to show that this statement follows also as a corollary from our
classification.

As shown by Gustavsson \cite{G2}, the fundamental identity (\ref{BLfund}) can
be thought of as a condition for the closure of the Lie algebra  defined from
an associative algebra of linear operators $\left[T^a,T^b,\cdot\right]$ on the
3-algebra. He then showed that any solution to the fundamental identity can be
written in the form $f_{abcd} = K_{pq} T_{ab}^p T_{cd}^q$, where $T_{ab}^p$ are
representation matrices of some Lie algebra, which was precisely our starting
point. For the ${\cal N}=8$ theory the constants have to be fully
antisymmetric, whereas for ${\cal N}=6$ only a weaker condition of antisymmetry
in $a$ and $c$ was required.  Recently similar discussion has appeared in  \cite{BL3} for $\cN=6$.

It is now a matter of simple examination of the solutions (\ref{su-solution})
and (\ref{sp-solution}) of the weaker conditions, to see that complete
antisymmetry can be achieved only for the $SU(2)\times SU(2)$ case.

\paragraph{Acknowledgments}
The authors thanks O. Bergman, D. Gaiotto and J. Maldacena for discussions.
 Both authors were supported by the
United States DOE Grant DE-FG02-90ER40542.
YT is also in part supported by the Carl and Toby Feinberg fellowship at
the Institute for Advanced Study.

\end{document}